\documentclass [12pt]{article}
\usepackage[cp1251]{inputenc}
\usepackage {amssymb}
\usepackage {amsmath}
\usepackage{epsfig}

\begin{document}

\begin{center}

{\Large\bf Calculation and analysis of solitary waves and kinks\\ in elastic tubes}%
\vskip 1cm

 I.B.Bakholdin. Keldysh Institute for Applied Mathematics.
125047. Miusskaya sq.4. Moscow. Russia. \textit{bakh@orc.ru}
\footnote{The work was supported by the Russian Fund for Basic
Research, grant 11-01-00034-a, and President Program of Support of
Leading Scientific Schools, grant NS-1303.2012.1.}

\end{center}

The paper is devoted to analysis of different models that describe
waves in fluid-filled and gas-filled elastic tubes and development
of methods of calculation and numerical analysis of solutions with
solitary waves and kinks for these models. Membrane model and
plate model are used for tube. Two types of solitary waves are
found. One-parametric families are stable and may be used as shock
structures. Null-parametric solitary waves are unstable. The
process of split of such solitary waves is investigated. It may
lead to appearance of solutions with kinks. Kink solutions are
null-parametric and stable. General theory of reversible shocks is
used for analysis of numerical solutions.

\section{Basic equations and formulae}
Solutions of equations of motion for elastic incompressible
cylindrical tube with fixed internal and external pressure
\cite{YbP} are analyzed here
\begin{equation}\left(R\sigma_1\frac{z'}{\lambda_1^2}\right)'-p_*rr'=\rho R
\ddot{z},\quad
\left(R\sigma_1\frac{r'}{\lambda_1^2}\right)'-\frac{\sigma_2}{\lambda_2}+p_*rz'=\rho
R \ddot{r}\label{baseeq}\end{equation}
$$\lambda_1=\sqrt{r'^2+z'^2},\quad \lambda_2=\frac{r}{R},\quad
\lambda_3=\frac{h}{H}$$
$$\sigma_i=\lambda_iW_{\lambda_i}-p$$
Here $'$ denotes time differentiation by variable $Z$, which is
Lagrange initial longitudinal spatial coordinate, $\dot{} $
denotes derivation by variable $t$ which is time. Thin membrane
model is used, unknowns  $z$ and $r$ determine surface of the tube
in cylindrical system of coordinates, $z$ axis of this system
coincides with central line of the tube. Parameter $p_*$ is
difference between internal and external pressure, unknown $p$ is
pressure in the material. Parameter $\rho$ is density of material
per unit of square so thickness of material which is denoted as
$h$ is not the parameter of these equations. Parameter $h$ will be
essential below in section \ref{last}. Here $\lambda_i$ are
principal stretches and $\sigma_i$ are principal Caushi stresses,
$W$ is stress function.

 It is assumed that in
the case of absence of any external forces
$$z=Z,\quad r=R,\quad h=H$$
Hence due to condition of incompressibility
$$\lambda_1\lambda_2\lambda_3=1$$
$$\sigma_i=\lambda_i\hat{W}_{\lambda_i}$$
$$\hat{W}=W(\lambda_1,\lambda_2,1/(\lambda_1,\lambda_2))$$
Stress function in this paper corresponds to Gent material
$$W=-\frac{1}{2}\mu
J_m\ln\left(1-\frac{\lambda_1^2+\lambda_2^2+\lambda_3^2-3}{J_m}\right)$$
In the case of investigastion of solitary waves (section
\ref{main}) initial data is taken in the form
$$z'=z_{\infty}'+\Delta z'(Z),\quad r=r_{\infty}+\Delta r(Z),\quad h=h_{\infty}+\Delta h(Z);\quad
\Delta r,\Delta z',\Delta h\to0,Z\to\infty$$
$$z(Z)=\int_0^Z z'(\zeta)d\zeta$$
Condition of equilibrium at infinity is posed
$$P_*=\frac{\hat{W}_{\lambda_2}(r_{\infty},z_{\infty}')}{r_{\infty}z_{\infty}'}$$

There are two equilibrium states at plus infinity and minus
infinity in the case of Riemann problem (section \ref{rim}).
Moreover, domains corresponding to other equilibrium states appear
as a result of solution of this problem. That is why analysis of
dispersion relation (section \ref{disp}) is fulfilled for some
arbitrary equilibrium state marked as zero state.

 Notations $\hat{W}_1=\hat{W}_{\lambda_1}$ and
so on will be used below.

\section{Analysis of dispersion relation}\label{disp}
In order to analyze  results of calculations examination of
dispersion relation \cite{YbI} is necessary. Dispersion relation
may be obtained after substitution of $z=z_l\exp(i(kZ-\omega)t)$,
$r=r_l\exp(i(kZ-\omega)t)$ into linearized version of equations
(\ref{baseeq}). Note that here physical frequency  and wavenumber
in Lagrange approach is used because all  graphs and calculations
correspond to this approach also. In Euler approach wavenumber
equals to $k/\lambda_1$. It is the physically observed wavenumber.
Equations are linearized  near some uniform state $z_0'$ and
$r_0$.

$$\omega\!=\!\pm\sqrt{\!\frac{\sigma_{10}}{\rho
z_{0}'^2}k^2\!+\!\frac{b\!\pm\!\!
\sqrt{b^2\!+\!4[(P^*r_0\!-\!\hat{W}_{210})^2\!-\!
(-\frac{\hat{W}_{220}}{R}\!+\!P^*z_0')(\frac{R\hat{W}_{10}}{z_0'}\!-\!R\hat{W}_{110})]k^2
} } {2\rho R} }$$
$$b=-\frac{R\hat{W}_{10}}{z_0'}k^2+R\hat{W}_{110}k^2+
\frac{\hat{W}_{220}}{R}-P^*z_0',\quad
$$

Let's introduce some notations

$$g=\hat{W}_{110}/(\rho R),\quad f=\sigma_{10}/(\rho z_0'^2),\quad U_l=\sqrt{g},\quad U_\tau=\sqrt{f},
\quad \omega_0=\sqrt{\frac{\frac{\hat{W}_{220}}{R}-P^*z_0'}{\rho
R}}$$ 




If $g>f$ then for sign $+$ before internal square root (plus
branch) $\omega/k\to U_l$ when $k\to +\infty$. Hence this branch
may be treated at infinity as longitudinal branch . For sign $-$
before internal square root (minus branch) $\omega/k\to U_\tau$
when $k\to +\infty$. Hence this branch may be treated at infinity
as transversal branch. If $g<f$ then for plus branch $\omega/k\to
U_\tau$ and for minus branch $\omega/k\to U_l$. Examples of graphs
of dispersive curves are presented in section \ref{main}
(fig.\ref{1a} and fig.\ref{4}).

For plus branch $+$
$\omega\to \omega_0$, $c\to +\infty$ when $k\to 0+$ and for minus
branch $\omega\to 0$, $c\to U_0$, $U_0$ is some finite value.

According to Petrovskii theorem \cite{kul} equations
(\ref{baseeq}) are non-correct if $U_{l}$ or $U_{\tau}$ is
imaginary value.  The first restriction of correctness is natural
for membrane model assumption $\sigma_1>0$, the second is also
natural for elasticity assumption that stretching of material case
resisting force.

The necessary stability conditions: $\omega_0\in \textsf{R}$,
$g>(P^*r_0-\hat{W}_{210})^2/(\rho R)$. Obvious the necessary
stability condition for uniform state is equilibrium condition
$$P_*=\frac{\hat{W}_2}{r_0 z_0'}$$
 But uniform state may be  unstable even if
equilibrium condition is fulfilled. Such states were really found
in analysis of solutions of Riemann problem (section \ref{rim}).

\section{Solutions of Boussinesq equations and analysis of numerical methods used in this paper}
This section is devoted mainly to development of numerical
methods. Solution of Boussinesq equations is treated as test
example for development of
methods of investigation. 

Boussinesq equations were derived for description of low-nonlinear
motions in the paper \cite{YbP}
\begin{equation}V_{\xi\xi}-c_1V_{\tau\tau}=V_{\xi\xi\xi\xi}+V^2_{\xi\xi}\label{bus}\end{equation}
Here $\tau$, $\xi$ and $V$ correspond to $t$, $Z$ and
$r-R_{\infty}$ after some stretching. We can also choose this
stretching so that $c_1=1$, because in the case $c_1<0$ according
to Petrovskiy theorem \cite{kul} this equation is not evolutionary
that means that it is not correct in the sense that Caushy problem
for such equation is not well-posed.

Standing solitary wave  solution for this equation takes the form
\cite{YbP}
\begin{equation}V_s(\xi)=\frac{3}{2}\textrm{sech}^2\left(\frac{\xi}{2}\right)\label{v0}\end{equation}
Note that this equation possesses also travelling solitary wave
solutions. Relation between standing solitary wave and travelling
solitary waves will be revealed below.

Linearized version of equations (\ref{bus}) was derived in
\cite{YbP}. Linearization was fulfilled in vicinity of solution
(\ref{v0}) in order to investigate stability of standing solitary
wave. Then method of Evans function was used to determine real
positive eigenvalue $s$ the square of  which seems to be the
rational number $s=3/16$. This fact permits to determine
eigenfunction analytically
$$B(\xi)=-\textrm{sech} \left(\frac{\xi}{2}\right)+2\textrm{sech}^3\left(\frac{\xi}{2}\right)$$
Hence solution (\ref{v0}) must be unstable.

Results of numerical calculations of (\ref{bus}) are described
below. 
The  purpose is
development of methods of numerical analysis which will be used in
section \ref{main} for more complicated equations.

Centered three-layer numerical scheme with second order accuracy
was used
\begin{eqnarray*}
\frac{V_{k+1}^n+V_{k-1}^n-2V_k^n}{\Delta
\xi^2}-\frac{Q_{k}^{n+1}-Q_{k}^{n-1}}{2\Delta \tau}
=\frac{V_{k+2}^n+V_{k-2}^n-4V_{k+1}^n-4V_{k-1}^n +6V_k^n}{\Delta
\xi^4}\\+\frac{(V^2)_{k+1}^n+(V^2)_{k-1}^n-2(V^2)_k^n}{\Delta
\xi^2},\quad \frac{V_k^{n+1}-V_k^{n-1}}{2\Delta
\tau}=Q_k^n\end{eqnarray*} Here $k$ and $n$ are spatial and time
indexes of numerical mesh, $\Delta\xi$ and $\Delta \tau$ are
spatial and time steps, $Q=V_\tau$.

Similar non-dissipative schemes were effectively used in various
problems of theory of nonlinear dispersive waves such as water
waves under elastic sheet waves  in composite material and
electronic magnetohydrodynamics \cite{mybook}. The other
appropriate scheme is two-layer Lax-Wendroff type scheme. For this
scheme approximation of spatial derivatives is the same as in
scheme given above but calculation of time derivatives is the same
as in Runge-Kutta method for ordinary differential equations. This
scheme is also scheme with second-order accuracy but it possesses
higher-order dissipation. Three-layer symmetrical scheme is more
preferable in non-dissipative problems and Lax-Wendroff scheme is
used for verification of results.  Additional dissipative terms
(viscous elasticity) may be included in both schemes if necessary
but it is made in different ways. Both schemes are conditionally
stable. Condition of stability may be obtained by spectral method
or by numerical experiment. For $\Delta \xi\to 0$ it typically
takes the form $\Delta\tau<c\Delta\xi^l$, $c=const$. For
three-layer method $l$ is usually equals to degree of grough
$\omega(k)$ for $k\to\infty$, here $\omega=\omega(k)$ is
dispersion relation of equations under consideration, or  to order
of highest spatial derivative. In all calculations in this paper
relation between time and spatial steps is far from the region of
linear numerical instability.

For non-scalar nonlinear equations slow nonlinear instability may
appear if three-layer scheme is used even if condition of linear
stability is fulfilled. Note that such instability was observed
for equations with low-order derivatives only. Calculations for
fixed-pressure equations showed that sometimes nonlinear numerical
instability really appears that is why when it happened
Lax-Wendroff scheme was used. Equations of fluid-filled model
considered in the last section are similar to gas dynamics
equations. Nonlinear instability of three-layer scheme for gas
dynamics is well-known so only Lax-Wendroff scheme is used.

Calculations are fulfilled for some bounded domain large enough to
avoid essential reflections. Rigid boundary conditions ($V=const$,
$r=const$, $z=const$) are posed.

Programs written on Fortran language are used.

No any graphs are presented in this section because the same
results were obtained  by calculation of full fixed-pressure
equations. Similar graphs are presented in the section below
(fig.\ref{1} and fig.\ref{2}).

Firstly initial data for numerical experiment was taken in the
form
$$V=V_s+\varepsilon B,\quad Q=\varepsilon sB$$

Just as expected for $\varepsilon>0$ on the first stage of the
evolution maximal value of $r$ grows with time and for
$\varepsilon<0$ it decreases.

In the case $\varepsilon>0$ the grough rapidly increases with time
and calculation stops due to overflow. This fact may be treated as
blowup. 



In the case $\varepsilon<0$ on the final stage of the process
standing solitary wave splits into pair of two travelling solitary
wave moving in opposite directions. No any other waves are
observed. We can make reverse of the process and treat standing
solitary wave as result of collision of two solitary waves moving
to each other. So standing solitary wave is special resonance
solution. Resonances between solitary waves are known for
two-dimensional models (Kadomtsev-Petviashvili equation) and for
models with high-order dispersion (generalized Korteweg-de Vries
equations and composite material equations) \cite{mybook}. But in
these cases one-parametric families of solutions exist. Some of
them are stable. The fact that standing solitary wave is the
result of interaction of two solitary waves suggests to make
conclusion that such split instability  of null-parametric
solitary waves will be observed for other models typically
\cite{new} and that for one-parametric resonance solitary waves
such split will be observed in some cases.



Equations (\ref{baseeq}) and (\ref{bus}) belong to class of
symmetrical equations for which waves moving with some fixed speed
are described by systems of symmetrical travelling wave equations.
Methods  for obtaining of symmetrical solitary waves developed in
\cite{mybook} may be applied for linearized versions of these
equations to obtain symmetrical eigenfunctions. Preliminary
estimates shows that corresponding shotting procedures in the case
of fourth-order equations require variation of one parameter if
eigenvalue is known from method of Evans function or two
parameters if eigenvalue is unknown. So calculations seems to be
not rather complicated. But investigation fulfilled below suggest
the method that do not requires calculation of eigenfunctions from
travelling wave equations for analysis of stability of solitary
waves.

Let initial data is
$$V=V_s,\quad Q=0$$
Numerical experiment  shows that after some time of standing
maximal amplitude groughs and then blowup happens. Moreover it was
opened that after some time of grough the difference
$\hat{B}=V(\tau_*)-V_s$ is similar to eigenfunction $B$. Initial
data
$$V=V_s+\varepsilon\hat{B},\quad
Q=\varepsilon\hat{Q}$$
cause increase and blowup for
$\varepsilon>0$ and decrease and split into two solitary waves for
$\varepsilon<0$.


 Results of last numerical experiments require some explanations.

It is well-known from mathematical physics that for linear models
with complete countable systems of eigenfunctions (it is typical
for problems posed for bounded domains) evolution after some time
is described mainly by eigenfunction with maximal positive
eigenvalue because any arbitrary disturbance typically contain
some finite component in expansion by eigenfunctions. But here we
have only one eigenfunction with positive value and problem for
infinite domain described by dynamical system of continuum type.
From formal point of view expansion of disturbance for arbitrary
initial data typically has zero component of this eigenfunction.
It seems that here eigenfunction in some sense is function of
maximal grough during the evolution and all disturbances that are
close to this function lead to evolution that is close to
evolution caused by disturbance proportional to eigenfunction.
This is explanation of observed phenomenons.

Type of evolution of numerical solution (growth or decrease) in
general case  depends on the difference between numerical solitary
wave and solitary wave taken as initial data. 
Other numerical method and other method to obtain initial data are
used in section below.  Initial decrease is observed for
$\varepsilon>0$ and blowup is observed for $\varepsilon<0$.

The other way to get approximation for eigenfunction without
analysis of solutions of ordinary differential equations is to
solve partial equations that are linearized near solitary wave.
Some initial disturbance is required  otherwise zero solution will
be obtained. Eigenfunction of maximal growth (or decrease) will
dominate after some time. But in general case for some special
types of disturbances eigenfunction may not appear. Only
oscillations will be observed. This is the main difference between
linear and nonlinear equations. In nonlinear case all waves are
related with each other. Hence any arbitrary disturbance leads to
appearing of all types of waves in nonlinear case.

Note also that in fact numerical experiments for determination of
eigenfunction of maximal growth are fulfilled for some bounded
domain. 

\section{Calculation of fixed-pressure
equations}\label{main}
 The main difference from the previous
section is that here standing solitary wave solution $r_s$, $z_s$
is numerical solution of travelling wave equations \cite{YbP}
obtained by the aid of some program written on Mathematica
language. This program was given by Y.B. Fu. This solution depends
from some parameters related with conditions at infinity. In the
limiting case standing solitary wave takes the form of two kinks
shifted to large distance one from another. According to method of
Evans function standing solitary wave solution is unstable
(exception is the kink case) but no analytic solution for
eigenfunction is available.

Three-layer non-dissipative centered numerical scheme with second
order of accuracy is given below
\begin{eqnarray*}
\frac{K_{k+1/2}^n\frac{z_{i+1}^n-z_{i}^n}{\Delta
Z}-K_{k-1/2}^n\frac{z_{i}^n-z_{i-1}^n}{\Delta Z}}{\Delta
Z}-p_*\frac{(r^2)_{k+1}^n-(r^2)_{k-1}^n}{4\Delta Z}=\rho
R\frac{\dot{z}_k^{n+1}-\dot{z}_k^{n-1}}{2\Delta t}\\
\frac{K_{k+1/2}^n\frac{r_{i+1}^n-r_{i}^n}{\Delta
Z}-K_{k-1/2}^n\frac{r_{i}^n-r_{i-1}^n}{\Delta Z}}{\Delta
Z}-(\hat{W}_{\lambda_2})_k^n-p_*r_k^n\frac{z_{k+1}^n-z_{k-1}^n}{2\Delta
Z}=\rho R\frac{\dot{r}_k^{n+1}-\dot{r}_k^{n-1}}{2\Delta t}\\
K_{k\pm 1/2}^n=((\hat{W}_{\lambda_1})_{k\pm 1}^n (\lambda_1)_{k\pm
1}^n+(\hat{W}_{\lambda_1})_{k}^n(\lambda_1)_{k}^n)/2 \\
\frac{z_k^{n+1}-z_k^{n-1}}{2\Delta t}=\dot{z}_k^n,
\quad\frac{r_k^{n+1}-r_k^{n-1}}{2\Delta t}=\dot{r}_k^n
\end{eqnarray*}
Instability is observed for long-time calculations.

 Lax-Wendroff
type scheme is more preferable here.
\begin{eqnarray*}
\frac{K_{k+1/2}^n\frac{z_{i+1}^n-z_{i}^n}{\Delta
Z}-K_{k-1/2}^n\frac{z_{i}^n-z_{i-1}^n}{\Delta Z}}{\Delta
Z}-p_*\frac{(r^2)_{k+1}^n-(r^2)_{k-1}^n}{4\Delta Z}=2\rho
R\frac{\dot{z}_k^{n+1/2}-\dot{z}_k^{n}}{\Delta t}\\
\frac{K_{k+1/2}^n\frac{r_{i+1}^n-r_{i}^n}{\Delta
Z}-K_{k-1/2}^n\frac{r_{i}^n-r_{i-1}^n}{\Delta Z}}{\Delta
Z}-(\hat{W}_{\lambda_2})_k^n-p_*r_k^n\frac{z_{k+1}^n-z_{k-1}^n}{2\Delta
Z}=2\rho R\frac{\dot{r}_k^{n+1/2}-\dot{r}_k^{n}}{\Delta t}\\
K_{k\pm 1/2}^n=((\hat{W}_{\lambda_1})_{k\pm 1}^n (\lambda_1)_{k\pm
1}^n+(\hat{W}_{\lambda_1})_{k}^n(\lambda_1)_{k}^n)/2 \\
2\frac{z_k^{n+1/2}-z_k^{n}}{\Delta t}=\dot{z}_k^n, \quad
2\frac{r_k^{n+1/2}-r_k^{n}}{\Delta t}=\dot{r}_k^n
\end{eqnarray*}

\begin{eqnarray*}
\frac{K_{k+1/2}^{n+1/2}\frac{z_{i+1}^{n+1/2}-z_{i}^{n+1/2}}{\Delta
Z}-K_{k-1/2}^{n+1/2}\frac{z_{i}^{n+1/2}-z_{i-1}^{n+1/2}}{\Delta
Z}}{\Delta
Z}-p_*\frac{(r^2)_{k+1}^{n+1/2}-(r^2)_{k-1}^{n+1/2}}{4\Delta Z}=\\
\rho
R\frac{\dot{z}_k^{n+1}-\dot{z}_k^{n}}{\Delta t}\\
\frac{K_{k+1/2}^{n+1/2}\frac{r_{i+1}^{n+1/2}-r_{i}^{n+1/2}}{\Delta
Z}-K_{k-1/2}^{n+1/2}\frac{r_{i}^{n+1/2}-r_{i-1}^{n+1/2}}{\Delta
Z}}{\Delta
Z}-(\hat{W}_{\lambda_2})_k^{n+1/2}-p_*r_k^{n+1/2}\frac{z_{k+1}^{n+1/2}-z_{k-1}^{n+1/2}}{2\Delta
Z}=\\\rho R\frac{\dot{r}_k^{n+1/2}-\dot{r}_k^{n}}{\Delta t}\\
K_{k\pm 1/2}^{n+1}=((\hat{W}_{\lambda_1})_{k\pm 1}^{n+1/2}
(\lambda_1)_{k\pm
1}^{n+1/2}+(\hat{W}_{\lambda_1})_{k}^{n+1/2}(\lambda_1)_{k}^{n+1/2})/2 \\
\frac{z_k^{n+1}-z_k^{n}}{\Delta t}=\dot{z}_k^{n+1/2}, \quad
\frac{r_k^{n+1}-r_k^{n}}{\Delta t}=\dot{r}_k^{n+1/2}
\end{eqnarray*}

\begin{figure}[!htb]
\centerline{\includegraphics[scale=0.6]{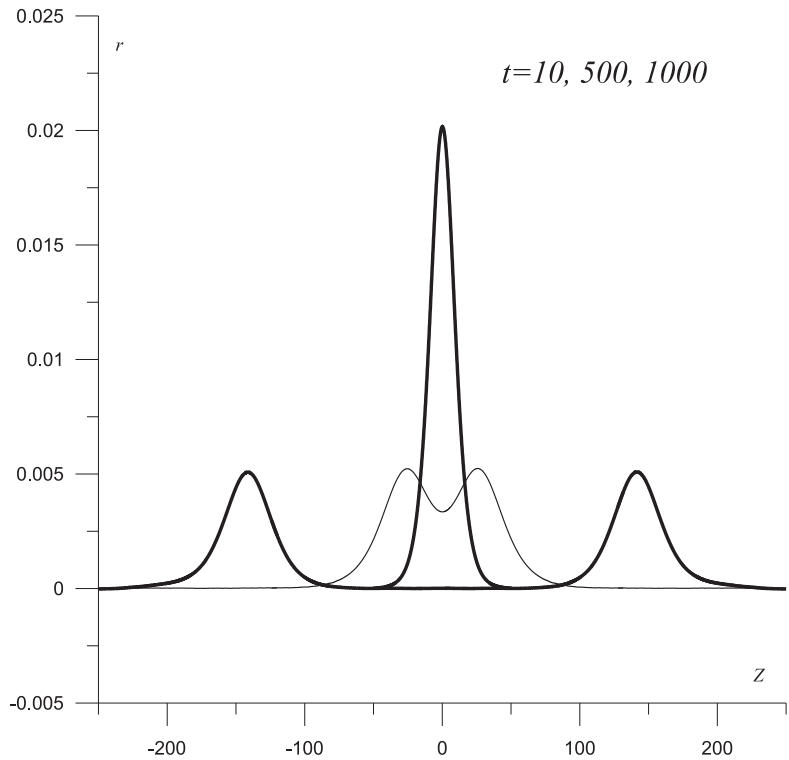}}
\caption{Small amplitudes: split}\label{1}
\end{figure}

Calculation with initial data
\begin{equation}r=r_s,\quad
z=z_s,\quad\dot{r}=0,\quad \dot{z}=0\label{o}\end{equation} after
some standing lead to decrease of maximal amplitude and split of
solitary wave. For the case of small amplitudes stationary
solitary wave splits into two solitary waves travelling to
opposite directions, fig.\ref{1}. So we have just the same type of
evolution that described in previous section.

\begin{figure}[!htb]
\centerline{\includegraphics[scale=0.6]{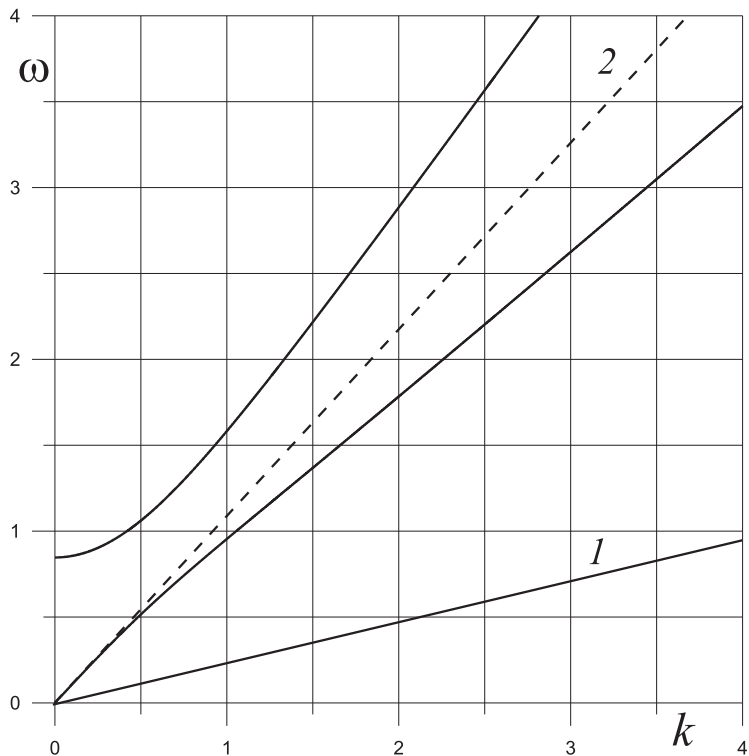}}
\caption{Small amplitude: dispersive curves}\label{1a}
\end{figure}

Dispersion curves are presented in fig.\ref{1a}. There is no
intersection between dispersive curves and line 1 corresponding to
speed of travelling solitary waves. Solitary wave of elevation as
exact solution may exist according to \cite{mybook} in principal
in this case but lower dispersive curve corresponds to the case of
negative dispersion while Boussinesq equation from the previous
section corresponds to the case of positif dispersion hence it may
describe some virtual waves. Line 2 corresponds to speed of linear
waves calculated according to results of calculations of
Boussinesq equation considered in previous section. According to
Boussinesq equation there is some constant ratio between speed of
travelling solitary waves appeared after split of standing
solitary wave and speed of linear waves. It equals approximately
0.43. This line is tangent to minus dispersive curve hence
Buossonesq equation describes these solitary waves very good.

The method developed in previous section was used to get solution
with increase of amplitude. The difference is that here
$\varepsilon<0$.

\begin{figure}[!htb]
\centerline{\includegraphics[scale=0.6]{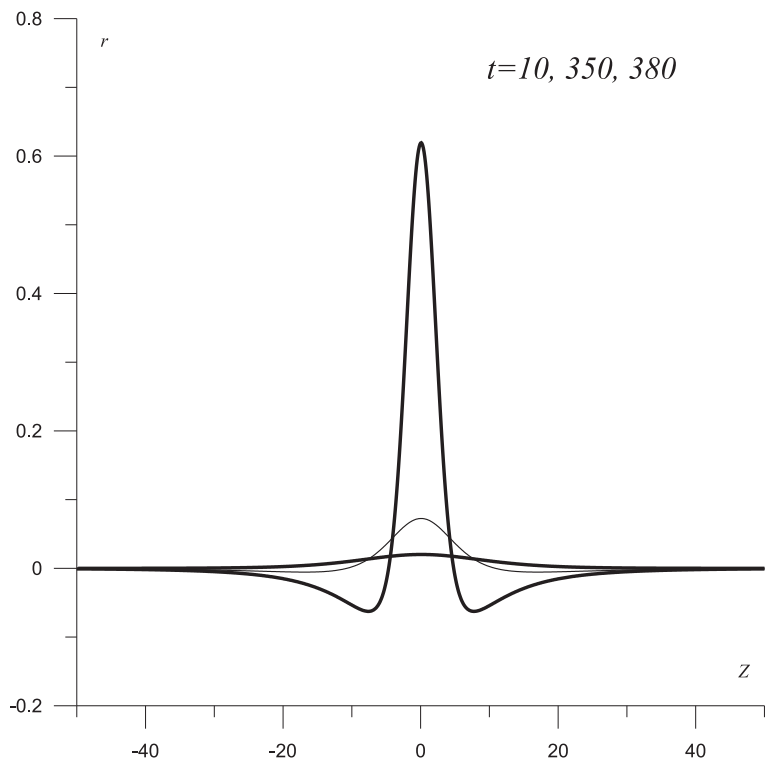}}
\caption{Small amplitudes: blowup}\label{2}
\end{figure}

Calculations with the initial data
$$z=z_s+\varepsilon(z(t_*)-z_s),\quad
r=r_s+\varepsilon(r(t_*)-r_s),\quad\dot{z}=\varepsilon\dot{z}(t_*),\quad\dot{r}=\varepsilon\dot{r}(t_*)
$$ give here increase of amplitude are for $\varepsilon<0$ and
decrease for $\varepsilon>0$. Here $z(t_*)$, $r(t_*)$ is solution
for initial data (\ref{o}) for some time $t_*$ when decrease is
already visible. In the case of small amplitudes  after some time
of increase calculations are stopped due to overflow, fig.\ref{2}.
Blowup may be. So evolution is  the same as described in previous
section.

\begin{figure}[!htb]
\centerline{\includegraphics[scale=0.6]{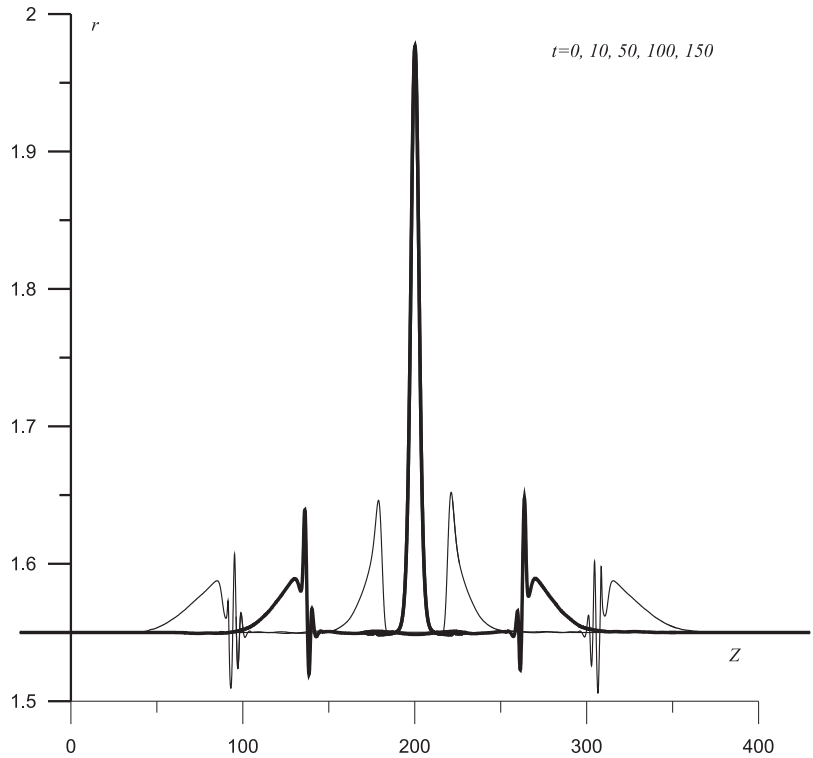}}
\caption{Case of moderate amplitude: initial decrease and
spli}\label{3}
\end{figure}

Behavior for the case of moderate amplitude differs from behavior
in the case of small amplitudes, fig.\ref{3}.

\begin{figure}[!htb]
\centerline{\includegraphics[scale=0.6]{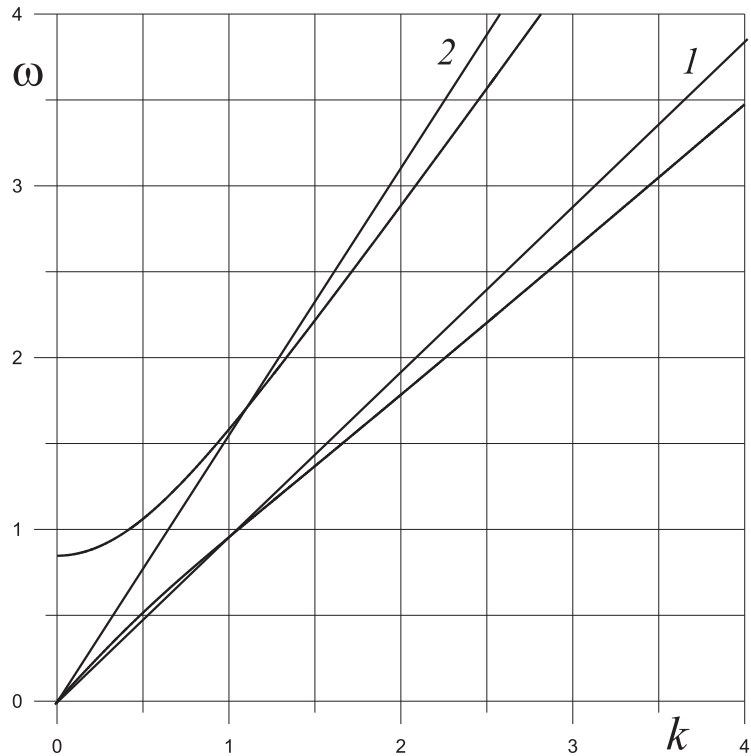}}
\caption{Case of moderate amplitude: dispersive curves}\label{4}
\end{figure}

In the case of decrease stationary solitary wave splits into two
waves also but this is not solitary waves. Amplitude of solitary
waves decreases due to dispersive effects. Short waves are issued.
Fig.\ref{4} shows dispersive curves for this case. Dispersion
relation is obtained after substitution
$z=z_l\exp(i(kZ-\omega)t)$, $r=r_l\exp(i(kZ-\omega)t)$ into
linearized version of equations (\ref{baseeq}). It was obtained in
\cite{YbI}. Note that here wavenumber in Lagrange approach is used
because all other graphs correspond to this approach also. In
Euler approach wavenumer equals to $k/\lambda_1$. It is the really
observed wavenumber.  Line 1 corresponds to speed of travelling
wave measured from numerical experiment. It intersects lower
dispersive curve for $k>0$. According to analysis of well-posed
tasks for numerical obtaining of solitary-wave solutions
\cite{mybook} there are no travelling solitary waves associated
with longitudinal branch for equations (\ref{baseeq}). Travelling
solitary waves mentioned in the previous section are only
approximate solutions here. If initial data will be taken in the
form of travelling solitary wave the tail of shot waves will
appear after some time and amplitude of this wave will decrease.
Nevertheless special resonance solitary wave exists. Note that by
this feature the model (\ref{baseeq}) is similar  to model of
composite material: classical solitary waves with monotonic
behavior at infinity do not exist but  resonance solitary wave
with such property was found \cite{bakhdif}.

\begin{figure}[!htb]
\centerline{\includegraphics[scale=0.6]{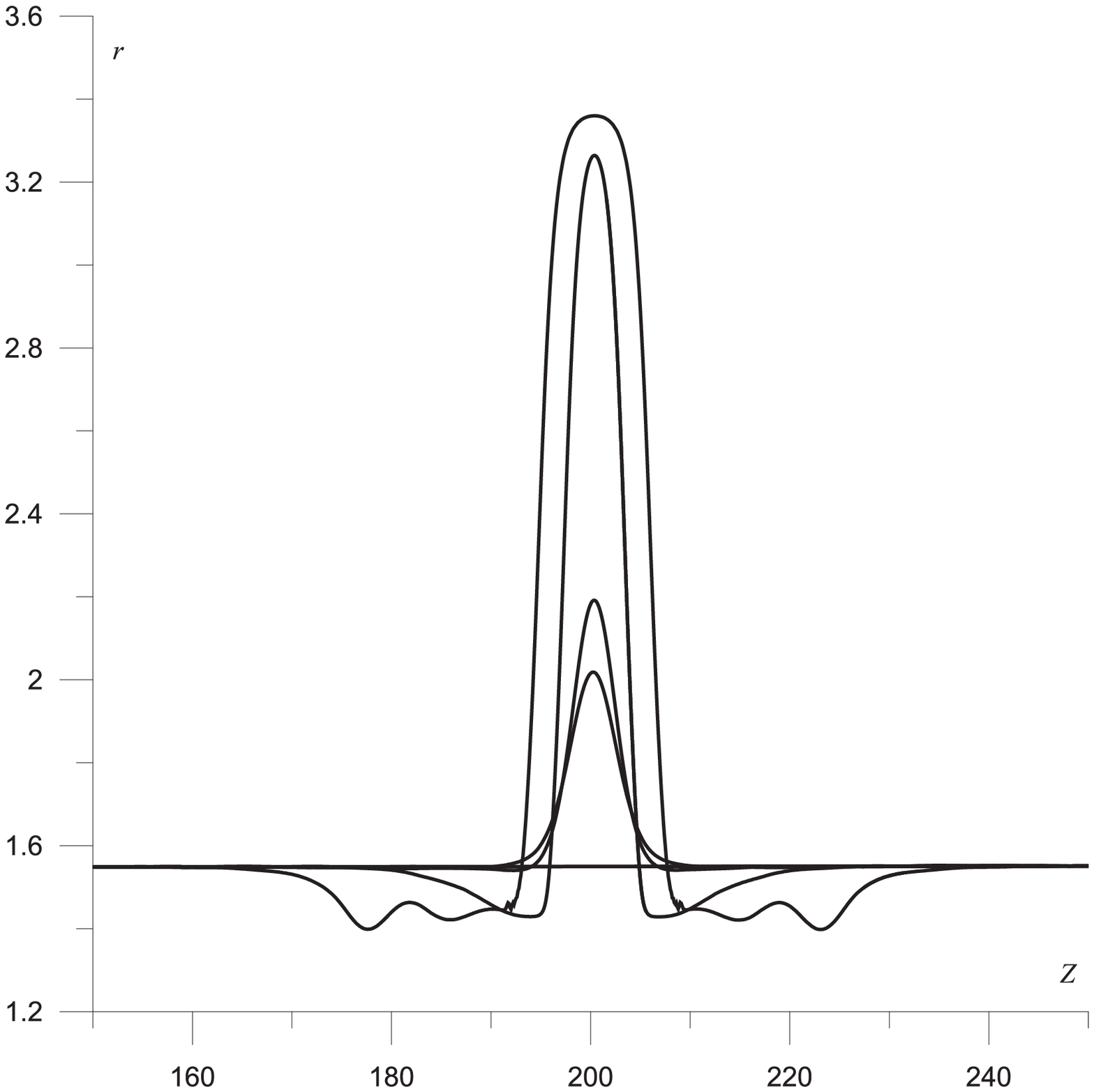}}
\caption{Case of moderate amplitude: initial growth and
stop}\label{5}
\end{figure}

\begin{figure}[!htb]
\centerline{\includegraphics[scale=0.6]{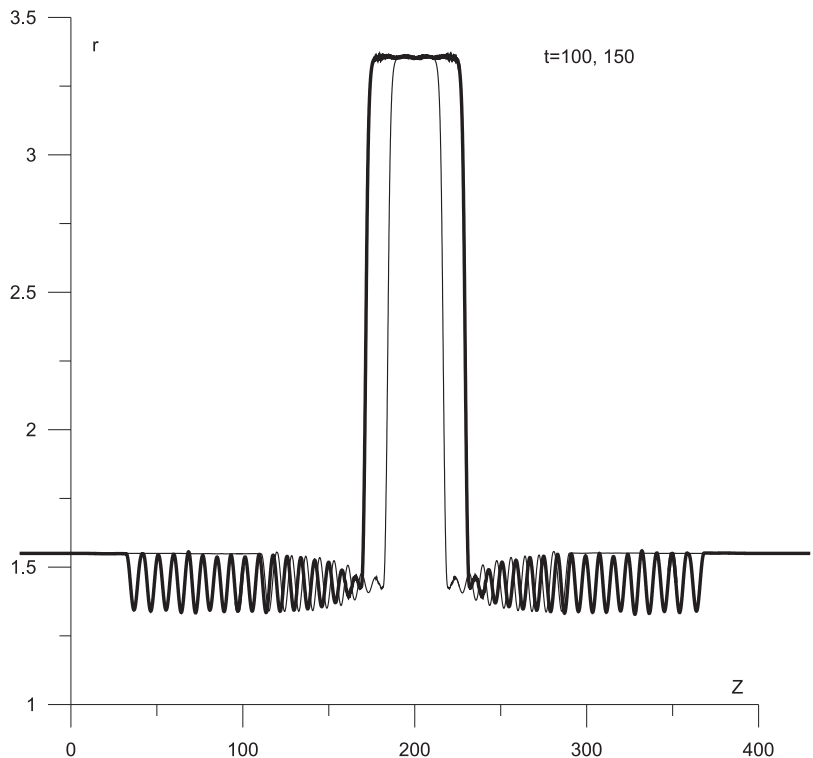}}
\caption{Self-similar solution}\label{6}
\end{figure}

In the case of increase for moderate amplitudes after some time
this increase stops (fig.\ref{5}) and envelop of graph of
resulting solution becomes self-similar (fig.\ref{6}). It depends
of $x/t$. This solutions contains two solitary-wave like shocks
(measured speed of the shock corresponds to line 2, there is
intersection, chaotic solution may appear after some time) and two
kinks moving in opposite directions. Analysis of graph for $z'$
shows that short longitudinal waves really appear here. Obviously
if stationary solitary wave will tend to combination of two kinks
then speed of kinks in this solution and amplitude of solitary
wave-like shock will tend to zero. Half of graph these solutions
(left or right side from axis of symmetry) coincides with the
graph of the special solution of Riemann problem, see section
below. Typical solution of Riemann problem contains two shocks or
simple waves and one kink. Here only kink and one shock are
observed. Hence maximal amplitude $r=r_{02}$ may be determined by
variation of amplitude of initial shock in solution of Riemann
problem.

\begin{figure}[!htb]
\centerline{\includegraphics[scale=0.6]{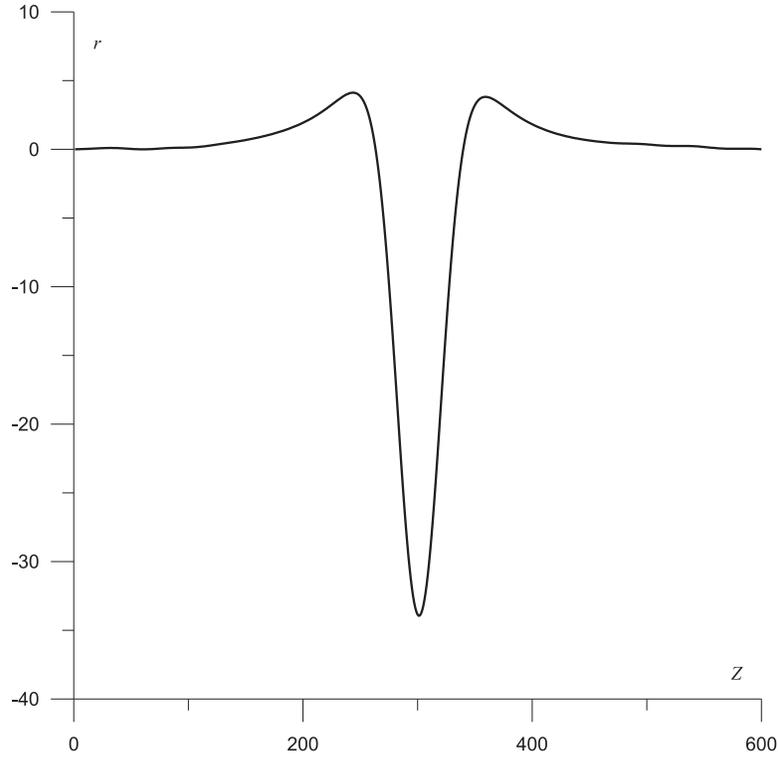}}
\caption{Moderate amplitude: eigenfunction}\label{7}
\end{figure}

For verification of results eigenfunction was calculated also as
solution of linearized equations, fig.\ref{7}. An interesting fact
is that eigenfunction as a result of calculations for a long
period of time is obtained not for all initial disturbances. For
some initial disturbances oscillations are observed for a long
time.

Though amplitude is moderate eigenfunction is similar to
eigenfunction of Boussinesq equations.

Calculation here are fulfilled  for $R=1$, $\rho=1$, $\mu=1$,
$J_m=30$, the same values correspond to graphs in sections below;
$z_{\infty}'=1.1$, $r_{\infty}=1.69$ (small amplitudes),
$r_{\infty}=1.55$ (moderate amplitudes).

\section{Arbitrary shock split  and kink solutions}\label{rim}
In order to verify stability of kink solution Riemann problem of
arbitrary shock split was solved. Initial data was taken in the
step-like form

$$r=r_{01}+(r_{02}-r_{01})\tanh((Z-Z_0)/L)/2,\quad \dot{r}=0,\quad r_{01}=r_{\infty}$$
$$z'=z_{01}'+(z_{02}'-z_{01}')\tanh((Z-Z_0)/L)/2,\quad \dot{z}=0,\quad z_{01}'=z_{\infty}'$$

Here $r_{02}$ is some arbitrary value and $z_{02}$ is calculated
by formula
$$P_*=\frac{\hat{W}_{\lambda_2}(r_{02},z_{02}')}{r_{02}z_{02}'}$$
This formula leads to problem of finding of the root of
fifth-order polynomial equation which is solved by Mathematika
package. Solution with property $\Delta=z_{02}'>z_{01}'$ is taken.
Note that there is two such solutions. The smaller value is taken
below. It means that branch bifurcating from zero for
$\Delta=z_{02}'-z_{01}'$ is used. The higher value leads after
some time of evolution to out of correctness calculations.
 Split into two or three shocks is observed. In the case of three shocks
one shock is kink mentioned above. But typically it is travelling
kink. Solution with standing kink is found by variation of
parameter $r_{02}$. Examples of calculations with positif and
negative speed of kink are shown below, fig.\ref{8}.

\begin{figure}[!htb]
\centerline{\includegraphics[scale=0.6]{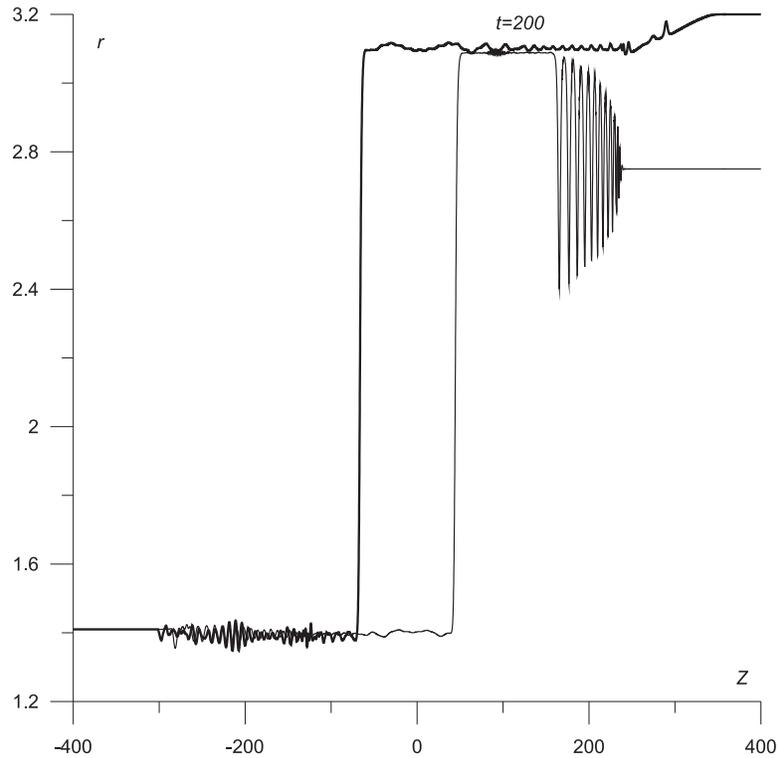}}
\caption{Solutions of Riemann problem close to stationary kink
solution. Initial shock was located at $Z=0$.\label{8}}
\end{figure}

Analysis of  shock problem solutions and dispersion relation for
uniform state of right side from kink shows that line tangent to
minus dispersive curve at the origin of coordinates do not
intersects it. Dispersion for the origin of coordinates is
positive so solitary-wave like shock seen on one of the graphs is
predicted by theory. Families of solitary waves of depression
tending to kink solution may exist also.

Examples of solutions for the case of large enough initial shock
are presented here. In the case of small amplitude no kink is
observed. In the case of moderate amplitude uniform state 2 is
unstable and calculation of Riemann problem leads to emergency
stop.

Now we can analyze another problem essential from the practical
point of view. Let we have initial data with two opposite steps:
$$r=r_{02},\quad -d<Z<d;\quad r=r_{01}\quad (-d>Z)\textrm{V}(Z>d)$$
Large scale approach is assumed here, $d$ is some distance.  Steps
are smoothed in reality. Standing kink solution in this problem
corresponds to some critical case. If the speed of left kink is
negative then such disturbance will increase it's area with time
and then the speed is positive the area of disturbance will
decrease.

The theory of non-dissipative shocks \cite{mybook}-\cite{bakhpmm4}
is applicable for this model and the investigations made above
were fulfilled by the aid of methods developed in this theory.
This theory is based on analysis of dispersion relation and
analysis of dimensions of invariant manifolds of solutions of
systems of equations of travelling waves numerical modelling of
invariant manifolds and direct calculations of partial equations.
So the same structures as in models described in
\cite{mybook}-\cite{bakhpmm4} exists. Extending non-local shocks
with solitary wave structure are observed then line $U=\omega/k$
does not intersect dispersive curve and shocks with stochastic
structure are observed in the
case of intersection. 

The equations with controlled pressure are Hamilton equations. But
there is no conservation of energy here. Hamiltonian here is not
energy. It seems that equations of of travelling waves for this
model are non-integrable.

\section{Derivation of simplified equations}
Simplified equations are required to describe centered simple
waves. In order to derive simplified equations let's introduce new
notation
$$\quad u=z'$$
Now equations takes the form
\begin{equation*}\left(R\hat{W_1}\frac{u}{\sqrt{u^2+r'^2}}\right)''-p_*(rr')'=\rho R
\ddot{u}\end{equation*}
$$\left(R\hat{W_1}(u,r,r')\frac{r'}{\sqrt{u^2+r'^2}}\right)'-\hat{W_2}(u,r,r')+p_*ru=\rho
R \ddot{r}$$
$$-\hat{W_2}(u,r,r')+p_*ru=0$$
Then let's take $u$ from second equation and put it to the first
equation and withdraw higher order derivatives. Resulting
equations takes the form
$$\Phi(r'',\ddot{r},..\textrm{other.second.order.terms}...;r.\textrm{coefficients.only}.)=0,\ u=\Gamma(r)$$
Here $\Phi$ and $\Gamma$ are some functions.

\section{Derivation of equations with bending resistance and
calculation}\label{last} For moderate values of $\varepsilon$
zones where $\sigma_1<0$ appear in the solution. This means
according to Petrovski theorem \cite{kul} that in this zones
equations (\ref{bus}) are non-correct. The other problem is that
on the boundary of correct and non-correct zone graphs $r(z)$ and
$r(Z)$ take the form of broken line. So we have discontinuity of
derivative here.

The model formally permits overlapping and breaking but analysis
showed that such phenomenons may appear only in the case of
compression and hence non-correctness.

One more problem is that there is no dispersion for short waves in
this model ($\omega/k\to const$ for $k\to +\infty$). It may cause
infinite growth of envelop.



Numerical scheme itself possesses it's own smoothing and
correctional properties so calculation continues some time after
non-correct zones appear but later shot-length oscillations arise
in non-correct zones and calculation stops due to increase of
these oscillations. Numerical solution in non-correct case is
usually only approximate solution. Usually there is convergence of
numerical scheme in correct case if some stability condition holds
and there is no convergence for any choose of relation between
time and spatial steps  in non-correct case. Hence non-correctness
may be revealed numerically also. Petrovskii theorem gives only
the necessary condition of correctness so then correctness must be
verified numerically.

Obviously  non-correctness is related to the fact that membrane
model can be used only for the stretched case. Plate model which
takes into account bending resistance works in stressed case also.
Bending resistance also obviously prevents the plate  from braking
that is from appearance of discontinuities for $r'$. Correction of
equations is described in section \ref{last}. Note that viscous
elasticity model may be used for correction also.

First of all let's note that the manner of derivation here is
differs sufficiently from the manner in which equations
(\ref{bus}) were derived. Derivation here is based on formulae of
linear theory of elasticity and converting of formulae of Euler
approach  into Lagrange form. The main goal is condemnation of
hypothesis made in previous section that implementation of bending
resistance may resolve problems appeared in calculations.

Let's start from well-known Germen-Lagrange formula for stable
state of thin unstretched and unstressed plate
$$p_l=\frac{Eh^3}{12(1-\sigma^2)}\eta_{xxxx}$$
Here $p_l$ is pressure acting on the plate, $\eta$ is vertical
displacement, $h$ is horizontal coordinate, $h$ is thickness, $E$
is Yung module, $\sigma$ is Poisson coefficient. Action of
internal elastic forces caused by bending of the plate may be
replaced by action of corresponding pressure.

For small stretch additional vertical force must be included in
left-right side of equations (\ref{base})

$$df_r=\frac{1}{3}\mu R r''''dZ$$

Cylindrical form of surface and incompressibility of material
$\sigma=1/2$, $E=3\mu$ are taken into account here.

Equation for $r$ takes the form
$$
-cr'''''+\left(R\sigma_1\frac{r'}{\lambda_1^2}\right)'-\frac{\sigma_2}{\lambda_2}+p_*rz'=\rho
R \ddot{r}\label{base}$$ Analysis of dispersion relation shows
that now equations  are  correct independently of whether
$\sigma_1>0$ or $\sigma_1<0$. Calculation with the same parameters
as in calculation described in previous section was fulfilled, $c$
was treated as a small parameter. No any problems are found. 


Solitary wave splits in the case of correction.

\section{Solution of equations for fluid-filled model}
Equations of equations filled by fluid are given below \cite{YbI}
\begin{equation*}\left(R\sigma_1\frac{z'}{\lambda_1^2}\right)'-Prr'=\rho R
\ddot{z},\quad
[[-cr'''']]+\left(R\sigma_1\frac{r'}{\lambda_1^2}\right)'-\frac{\sigma_2}{\lambda_2}+Prz'=\rho
R \ddot{r}\end{equation*}
$$\dot{r}z'-r'\dot{z}+vr'+\frac{1}{2}rv'=0,\quad\rho_f(\dot{v}z'-v'\dot{z}+vv')+P'=0$$
Here $v$ is fluid velocity, $\rho_f$ is fluid density.
Correctional term discussed in previous section is included here
in double square brackets.

Dispersion relation for these equations was derived in \cite{YbI}.
The main difference from the case of fixed-pressure equations is
that both branches intersect in point $\omega=0$, $k=0$.

It is not clear how to calculate these equations. The one way is
replace fluid gas by low-compressible gas. The required
modification of  hydrodynamic part of equations is given below

$$(\dot{\rho_f}z'-\rho_f'\dot{z})r^2+2\rho_f r(\dot{r}z'-r'\dot{z})+(\rho_f v r^2)'=0$$
$$P=P(\rho_f)$$
Equation of conservation of mass is modified here. Equation of
state is added. The problem is wheather dispersion can prevent
breaking of waves in gas if $c=0$.

 There are three dispersive curves for these
equations.

The other way is to eliminate pressure and get three equations

\begin{eqnarray*}
\left[\rho_fz'-\left(\frac{\frac{r'\rho
R}{rz'^2}}{1+\frac{r'^2}{z'^2}}\right)'\right]\dot{v}-
\left[\left(\frac{\frac{r'\rho
R}{rz'^2}}{1+\frac{r'^2}{z'^2}}\right)+\left(\frac{\frac{\rho
R}{2z'^2}}{1+\frac{r'^2}{z'^2}}\right)'\right]\dot{v}'-
\left(\frac{\frac{\rho
R}{2z'^2}}{1+\frac{r'^2}{z'^2}}\right)\dot{v}''=\\
\rho_f(v'\dot{z}-v v')-[[c r''''/(r z')]']+\quad\\
\left[\frac{\frac{(1-r')\left(R\sigma_1\frac{r'}{\lambda_1^2}\right)'
-\frac{\sigma_2}{\lambda_2}}{r z'}-\frac{\rho
R}{rz'^2}(\dot{z}-v)\left(\frac{z'\dot{z}-v
r'-\frac{1}{2}rv'}{z'}\right)'- \frac{\rho R}{rz'^3}(r'\dot{z}-v
r'-\frac{1}{2}rv')(\frac{1}{2}v'-\dot{z}')}{1+\frac{r'^2}{z'^2}}\right]'\\
\end{eqnarray*}
$$\dot{z}=q$$
$$
\rho R\dot{q}=\left(R\sigma_1\frac{z'}{\lambda_1^2}\right)'-Prr'
$$
\begin{eqnarray*}
P=[[c r''''/(r
z')]]+\frac{-1}{1+\frac{r'^2}{z'^2}}\left[\frac{r'\rho
R}{rz'^2}\dot{v}+\frac{\rho R}{2z'^2}\dot{v}'\right.\\
\frac{(1-r')\left(R\sigma_1\frac{r'}{\lambda_1^2}\right)'
-\frac{\sigma_2}{\lambda_2}}{r z'}-\frac{\rho
R}{rz'^2}(\dot{z}-v)\left(\frac{z'\dot{z}-v
r'-\frac{1}{2}rv'}{z'}\right)'\\\left.- \frac{\rho
R}{rz'^3}(r'\dot{z}-v
r'-\frac{1}{2}rv')(\frac{1}{2}v'-\dot{z}')\right]
\end{eqnarray*}

These equations are solved by Lax-Wendroff type numerical scheme.
Due to combined time-spatial derivatives this scheme is implicit.
System of implicit equations is solved by method of iterations.
Example of calculation of Riemann problem is given below,
fig.\ref{9}, $v|_{t=0}=0$, $\rho_f=1$. Kink and solitary wave-like
shocks are clearly seen.

For solitary waves used for shock structures for some given values
of physical parameters  and phase speed $U$ typically one solution
exist. Such one-parametric ($U$ is arbitrary parameter) families
of solitary waves in Hamilton systems are typically stable. Here
for some special choose of $v$ at infinity these solitary waves
may be stationary. There are also other null-parametric stationary
solitary waves \cite{ilib} that are similar to waves investigated
for the case of controlled pressure. They exist only for $v=0$.
Null-parametric solitary waves in Hamilton systems are typically
unstable because if energy of solitary wave is slightly decreased
there is no way to return it back. Instability of this waves was
verified in \cite{ilib} by Evans method.

\begin{figure}[!htb]
\centerline{\includegraphics[scale=0.6]{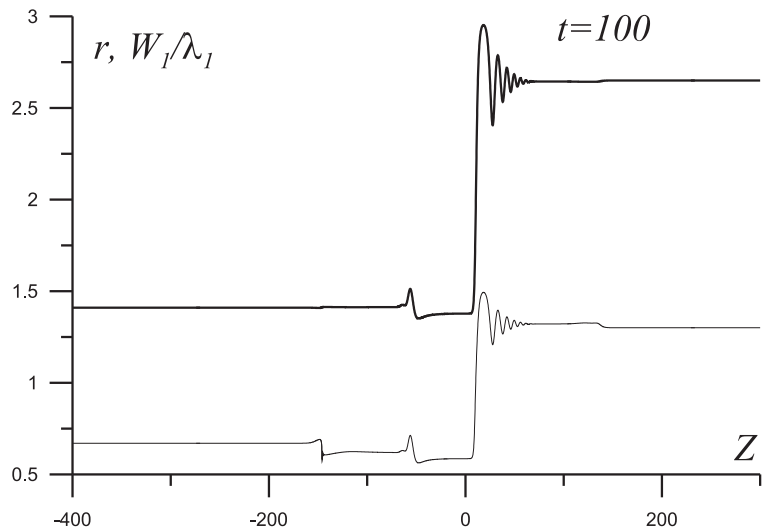}}
\caption{Fluid-filled model: Riemann problem \label{9}}
\end{figure}

 The fluid-filled model is the model with conservation of energy. It
leads in the case $c=0$ to integrable system of travelling wave
equations and only classical solitary waves and kinks are
expected.

Note that the model under consideration due to hydraulic
approximation implies   slow variation of $r(Z)$ hence
simplifications may be fulfilled such that no space-time
derivatives will be. But such simplification obviously leads to
decrease in accuracy and appearance of non-physical effects. For
low-pressure case further simplification may be fulfilled and
classical or generalized (if bending resistance is included)
Korteweg-de Vries equations may be derived. Detailed analysis of
solutions of generalized Korteweg-de Vries equation is fulfilled
in \cite{mybook} and \cite{bakhpmm4}

\end{document}